# Highly tunable repetition-rate multiplication of mode-locked lasers using all-fibre harmonic injection locking

Chan-Gi Jeon, Shuangyou Zhang, Junho Shin, and Jungwon Kim*

*School of Mechanical and Aerospace Engineering, Korea Advanced Institute of Science and Technology (KAIST), Daejeon 34141, Korea*

*e-mail: jungwon.kim@kaist.ac.kr

**Higher repetition-rate optical pulse trains have been desired for various applications such as high-bit-rate optical communication, photonic analogue-to-digital conversion, and multi-photon imaging. Generation of multi GHz and higher repetition-rate optical pulse trains directly from mode-locked oscillators is often challenging. As an alternative, harmonic injection locking can be applied for extra-cavity repetition-rate multiplication (RRM). Here we have investigated the operation conditions and achievable performances of all-fibre, highly tunable harmonic injection locking-based pulse RRM. We show that, with slight tuning of slave laser length, highly tunable RRM is possible from a multiplication factor of 2 to >100. The resulting maximum SMSR is 41 dB when multiplied by a factor of two. We further characterize the noise properties of the multiplied signal in terms of phase noise and relative intensity noise. The resulting absolute rms timing jitter of the multiplied signal is in the range of 20 fs to 60 fs (10 kHz – 1 MHz) for different multiplication factors. With its high tunability, simple and robust all-fibre implementation, and low excess noise, the demonstrated RRM system may find diverse applications in microwave photonics, optical communications, photonic analogue-to-digital conversion, and clock distribution networks.**

Higher repetition-rate optical pulse trains have been desired for various applications such as high-bit-rate optical communication[1,2], optical clock distribution[3,4], photonic analogue-to-digital conversion[5,6], calibration of astronomical spectrograph[7], multi-photon imaging[8] and low noise microwave generation[9,10]. Generation of high (e.g., multi GHz and higher) repetition-rate optical pulse trains directly from mode-locked oscillators is often challenging. Fundamental mode-locking at GHz repetition-rate requires high pumping power and short cavity length[11-13] and is generally more susceptible to Q-switched mode-locking[14]. To raise the repetition-rate beyond fundamental repetition-rate, harmonic mode-locking has been used for multiple pulse circulation in the cavity. This is achieved by various techniques such as active phase modulation[15,16], nonlinear polarization rotation with dispersive radiation in long cavity[17,18] or real saturable absorbers[19,20]. However, it requires sophisticated cavity design for intended harmonic mode-locking condition, and is usually sensitive to the operation environment. It also suffers from unwanted supermodes[21,22] and often requires active stabilization to suppress them[23-25].

On the other hand, extra-cavity repetition-rate multiplication (RRM) can be an effective solution for achieving multi-GHz repetition-rate from standard ~100-MHz mode-locked lasers. For this, several different methods, such as Fabry-Pérot (F-P) etalons[26-29], Mach-Zehnder interferometers[30-32] and sub-ring fibre resonators[33-35], have been demonstrated. Precisely fabricated F-P cavity filters frequency modes by more than GHz free-spectral-range and this mode selecting can be stabilized by well-developed stabilization techniques[36,37]. However, it is generally costly and alignment-sensitive, and has a fixed input-output frequency relationship. A 2-by-2 fibre Mach-Zehnder interferometer can interleave pulse train itself with serial steps of duplication by factor of $2^n$. Here each stage requires individual delay tuning and the path length error can be easily accumulated. Hence, only few stages of cascading are available. All-pass sub-ring fibre resonator is advantageous for pulse energy conservation, but can achieve only few times of multiplication factor. In common, these techniques cannot avoid intrinsic power loss proportional to the multiplication ratio.

As an alternative, injection locking can be applied for the extra-cavity RRM system. Laser injection locking has been mainly used for amplification[38] and duplication[39-42] of low-noise laser oscillator. Carefully designed slave laser can inherit characteristics of the master laser signal with fairly low power of injection. Among several injection locking-based RRM methods demonstrated so far[43-46], harmonic injection locking is an effective way to scale up the factor of multiplication by the Vernier effect[44-46], which recently resulted in 25 dB maximum side mode suppression ratio (SMSR) when multiplied by a factor of 25 (from 40 MHz to 1 GHz)[46].

In this paper, we have investigated the operation conditions and resulting performances of a harmonic injection locking-based all-fibre RRM system. Unlike the previous work that used carrier-envelope offset frequency ($f_{ceo}$) locked slave laser condition[46], we investigated the performance under free-running operation of slaver laser without offset frequency synchronization. Despite no offset frequency control, long-term stable (e.g., >12 hours continuous operation) RRM was possible. We show that, with slight tuning of slave laser cavity length, highly tunable RRM is possible from a multiplication factor of 2 to >100. The resulting maximum SMSR is 41 dB when multiplied by a factor of two. We further characterized the noise properties of the multiplied signal in terms of phase noise and relative intensity noise (RIN). The resulting absolute rms timing jitter of the multiplied signal is in the range of 20 fs to 60 fs (integration bandwidth: 10 kHz – 1 MHz) for different multiplication factor and SMSR conditions. With its versatile operation and high frequency tunability, simple and robust all-fibre implementation, and fairly low excess noise, the demonstrated RRM system may find diverse applications in microwave photonics, optical communications, optical sampling and photonic analogue-to-digital conversion, and clock distribution network systems.

**Results**

**Harmonic injection locking-based RRM.** Laser injection locking can be regarded as regenerative amplification of a master signal through a slave oscillator. In particular, the multimode interaction between optical frequency combs is described as a group of typical injection locking[47] and theoretically explained using the perturbation theory[39,40] by regarding the weak injection pulses from the master laser as perturbations on slaver laser mode-locking solution. When the mode spacing of two combs, i.e., the repetition-rate frequencies of master and slave lasers, is set to the integer ratio with least common multiple, harmonic injection locking is achieved (Fig. 1**a**). Here the injection-locked modes should be dominant enough to suppress the unwanted oscillation of the slave laser under proper operating condition.

The experimental system under test is shown in Fig. 1**b**. The master laser, a nonlinear polarization rotation-based mode-locked Erbium-fibre laser, has a 78.43 MHz repetition-rate. A circulator is used for both guiding injection of master signal and emitting the final repetition-rate multiplied output signal. The slave laser is a linear-cavity, semiconductor saturable absorber (SESAM)-mode-locked soliton Er-fibre laser with 14 cm length tunability by moving an end mirror mounted on a translation stage (see Methods for more information).

To characterize the major optical and radio-frequency (RF) spectral properties of the injection-locking-based all-fibre RRM system, we first focused on the case of multiplication factor of 13 (i.e., multiplication from 78.43 MHz to 1.02 GHz) by setting the slave laser repetition-rate at 72.84 MHz. Figure 2 shows the main characteristics of the measured spectra. The measured optical spectra are shown in Fig. 2**a**. The mode-locking spectrum of the slave laser (curve 1 of Fig. 2**a**) is centred at 1561 nm with 7 nm full-width-at-half-maximum (FWHM) bandwidth, which is much narrower than that of the master laser (curve 2 of Fig. 2**a**). To match the optical spectra between two lasers for more effective injection, optically filtered master signal (curve 3 of Fig. 2**a**) is used for the slaver laser input. Under proper filtering condition, efficient injection locking could be achieved (curve 4 of Fig. 2**a**). Note that the final output spectrum shows fringe-like patterns because this system is operated by two free-running lasers, where non-zero difference between carrier-envelope offset frequencies exists. When the repetition-rate multiplication factors are $M$ and $N$ for master oscillator (with free-spectral range of $f_M$) and slave oscillator (with free-spectral range of $f_S$), respectively, the resulting multiplied repetition-rate is $f_{output} = Mf_M$. When the carrier-envelope offset frequencies are different between master and slave oscillators, there should be a slight frequency detuning, $\varepsilon = |\, f_S - Mf_M/N\,|$ to achieve effective injection locking. From this frequency detuning, the fringe wavelength separation can be derived as $\Delta\lambda_{fringe} = \frac{\lambda_c^2}{c}\frac{f_S^2}{M\varepsilon}$, where $\lambda_c$ is the centre wavelength and c is the speed of light, when $N = M+1$. This agrees well with the measured optical spectrum (Curve 4 in Fig. 2**a**; $\varepsilon$ = 1.8 kHz leads to 1.9 nm fringe separation). Nevertheless, a sufficiently small error of the repetition-rate ratio within the injection locking range allows enough frequency modes to form the desired RRM condition. Note that, even without any $f_{ceo}$ control between the two lasers, the found RRM condition has been maintained for more than 12 hours.

Figure 2**b** shows the measured RF spectrum of the repetition-rate multiplied output signal. As expected, a strong injection-locked frequency mode is obtained at 1.02 GHz with much weaker side modes coexisting. Here, SMSR, the RF power ratio of the main mode to the most prominent side mode, is the typical index for evaluating the RRM quality. Proper tuning of injection ratio (i.e., the ratio of master injection power and emission power of slave laser) and optical filtering are required to maximize the SMSR (see Methods section). The maximum SMSR for RRM factor of 13 is measured to be 32 dB, which is comparable to transmission function of a single-pass FP cavity with finesse of ~300[27-29]. Figure 2**c** shows the time-domain waveform of the multiplied pulse train, measured by a 33-GHz real-time oscilloscope (Keysight, MSOV334A). As expected from the RRM factor of 13, the pulse train waveform shows an amplitude modulation with a period of 13 pulses. The measured maximum amplitude modulation depth of 5.6% agrees well with the calculated result from the measured SMSR of 32 dB.

We also assessed the long-term frequency drift between the maser and slave oscillators when injection locked. Figure 2**d** shows the long-term relative frequency difference result at 1 GHz. Note that, for this measurement, a 250-MHz master laser (*M*=4) and a 76.9 MHz slave laser (*N*=13) is used, which results in 1-GHz injection-locked output pulse train. One can see that the injection locking can be maintained over 12 hours, and when we intentionally break the injection, free-running operation between two lasers are clearly visible.

**Tunability in multiplication factor.** The main advantage of this harmonic injection locking-based RRM system is a highly tunable operation with Vernier effect. Note that the repetition-rate multiplication happens when a pair of co-prime integers $M$ and $N$ satisfies the following relationship, $Mf_M = f_{output} \approx Nf_S$, where $f_M$ and $f_S$ are the repetition-rates of the master and slave oscillators, respectively, and $f_{output}$ is the resulting multiplied repetition-rate. As a result, by tuning $f_S$ for a given $f_M$, numerous combinations of co-prime integer pair ($M$, $N$) exist, which enables high tunability in the master laser multiplication factor $M$. Figure 3 shows an example of available multiplication factors $M$ versus the repetition-rate of the slave laser (in the range of 51 MHz – 78 MHz) when the master laser repetition-rate $f_M$ is fixed at 78.43 MHz. Among different combinations of $M$ and $N$, we explore three cases more in detail.

First, for case **a** in Fig. 3, $N$ is set to ($M$+1) so that all $M$ factors can be continuously swept from 13 (1.02 GHz) to 128 (10.04 GHz) by tuning the slave oscillator end mirror by a 14-cm-long translation stage. Among them, measured RF spectra for four representative cases, ($M$,$N$)=(13,14), (27,28), (51,52) and (102,103), are shown in Fig. 4**a**. Case **b** in Fig. 3 indicates different $N$ values can achieve the same multiplication factor $M$. In Fig. 4**b**, three different cases of slave oscillator conditions ($N$ = 52, 53 and 55) are shown for $M$=51 multiplication (4 GHz). The interesting finding here is that, while the RF spectra have different shapes, they all have similar SMSR values of ~20 dB. In time domain, these three cases have different pattern of pulse-to-pulse waveform but similar amount of modulation depth. Finally, case **c** in Fig. 3 indicates low (e.g., <10) multiplication factor conditions. To realize these low M conditions, much lower repetition rate of slave oscillator is required, hence we added more SMF-28 fibre to the slave cavity. In Fig. 4**c**, the number of side modes is decreased and injection locked modes become more dominant. Correspondingly the SMSR is increased up to 41 dB at the multiplication factor of 2 and 3. Figure 5 shows the measured SMSR versus RRM factor of our system in comparison with previous extra-cavity RRM results[26-32,34,35,43,46]. The SMSR results are similar to or superior to the previous approaches up to RRM factor of ~30, and further shows higher tunability up to RRM factor of >100. Note that, even though $f_{ceo}$'s of the two lasers are not locked, the SMSRs are similar to the previously reported result with $f_{ceo}$ difference locking[46]. In order to check the impact of $f_{ceo}$ lock, we also implemented and tested the $f_{ceo}$ difference locking mechanism shown in ref. 46, but could not find clear difference or improvement in SMSRs.

We also tested the RRM performance when different master oscillators are employed for the same slave oscillator (with ~73 MHz repetition-rate): one is a home-built figure-9 laser at 35.7 MHz repetition rate[48] and the other is a commercial mode-locked oscillator at 250 MHz repetition rate (MenloSystems GmbH, FC1500-250-ULN). The harmonic injection locking worked well with easy initiation for all three master oscillators with vastly different repetition rates (35.7 MHz, 78.4 MHz and 250 MHz in this work). Figure 6 shows the collection of measured SMSRs for three different master oscillators with various multiplication factor conditions. For the 78.4 MHz master laser, the SMSR decreases from 41 dB (*M*=2) to 12 dB (*M*=128) with decent >30 dB SMSR kept up to the multiplication factor of ~20. Note that a 10-GHz photodiode is used for the RF spectrum measurements, which limited the measured SMSRs in Fig. 6 only up to multiplication factor of ~40 for the 250-MHz master oscillators. The actual RRM can happen beyond this RRM range. These results show that a single slave laser cavity (with length tuning mechanism such as an end mirror on a translation stage) can be used for various master lasers with vastly different repetition-rates.

**Phase noise and intensity noise of repetition-rate multiplied signals.** Phase noise and intensity noise power spectral densities (PSDs) of the multiplied pulses are characterized. First, four RRM conditions (where SMSR is maximized for each case) at 1.02 GHz, 2.04 GHz, 4.08 GHz and 8.16 GHz ($M$=13, 26, 52 and 104, respectively) are investigated. The phase noise PSDs are measured by a signal source analyser (Rohde & Schwarz, FSWP) at the same carrier frequency at 8.16 GHz. As shown in Fig. 7, the phase noise PSD shapes and levels are similar for any RRM factors. The integrated absolute timing jitters are less than 379 fs (62 fs) with maximum additive jitter of 316 fs (62 fs) when integrated from 10 Hz (10 kHz) to 2 MHz Fourier frequency. One interesting finding is that, as the RRM factor increases, the peak at ~240 kHz decreases. As a result, higher RRM factor leads to lower integrated timing jitter: when M is set to 102 (8.16 GHz), the integrated rms timing jitter decreases down to 19.6 fs. Note that the demonstrated high frequency timing jitter in the range of 20-60 fs is comparable to or lower than the time resolution of the currently available high-performance, high-speed oscilloscopes. The relative intensity noise (RIN) is measured by FFT analyser (Stanford Research Systems, SR770) and RF spectrum analyser (Agilent, E4411B) for <100 kHz and >100 kHz Fourier frequency, respectively. As the phase noise results (Fig. 7) were, the measured RIN PSDs (Fig. 8) show similar shapes and levels regardless of the RRM factor. The integrated RIN is less than 0.2 % when integrated from 10 Hz to 2 MHz Fourier frequency.

**Discussion**

We have investigated the operation conditions and achievable performances of all-fibre, highly tunable harmonic injection locking-based pulse RRM system. The RRM factor up to 128 (>10 GHz) from 78.43 MHz is demonstrated. The maximum SMSR is 41 dB for $M$=2 and the SMSR monotonically decreases down to 12 dB for $M$=128. Compared to other existing RRM methods such as Mach-Zehnder interferomter[30-32] and all-pass resonators[33-35], it has much higher multiplication tunability with similar SMSR performances. Moreover, the demonstrated SMSR is higher than that of single-pass cavity-filtering methods with F-P cavity with finesse ~ 300[27-29] for overall tested range. Only double-pass cavity-filtering methods[26,28,49], which require sophisticated locking electronics, can have much higher SMSRs than the demonstrated injection locking based RRM results. The baseband phase noise and intensity noise are also characterized, which shows similar levels regardless of the multiplication factor. The absolute high-frequency (e.g., >10 kHz Fourier frequency) timing jitter can be as low as ~20 fs. With its high tunability, simple and robust all-fibre implementation, and low excess noise, the demonstrated RRM system may find diverse applications in microwave photonics, optical communications, optical sampling, photonic analogue-to-digital conversion, and clock distribution network systems.

**Methods**

**Fibre mode-locked lasers for master and slave oscillators.** A nonlinear polarization rotation (NPR) mode-locked Erbium fibre laser of 78.43 MHz repetition-rate in a sigma-cavity is used for the main master laser. Under near zero net cavity dispersion (~0.002 ps$^2$) for stretched-pulse mode-locking regime, optical spectrum has more than 50 nm FWHM bandwidth centred at 1580 nm. The slave laser is a semiconductor saturable absorber mirror (SESAM)-based soliton mode-locked laser in a linear cavity with negative net cavity dispersion (~ -0.017 ps$^2$). This laser includes a short free space section with end mirror mounted on a 14-cm-long translation stage for cavity length tuning.

**Experiment conditions and outputs.** When a 90:10 coupler is used for the slave laser, its output power before injection locking is ~1.1 mW. The injection-locked output power is in a narrow range of 1.1 mW – 1.24 mW even when the input injection power is changed in a wide range of 8 mW – 24 mW. In addition, the output power varied less than 0.2 dB for all multiplication factor conditions when the input injection power is maintained. The output pulsewidth is ~1.5 ps, which is almost independent on the input pulsewidth condition (e.g., injecting 100 fs or 1.7 ps resulted in a similar output pulsewidth). One thing to note is that, despite the duration of the main pulse is maintained, the temporal positions of small pre- and post-pulses depend on the fringe spacing and shape of the output optical spectra (Fig. 2**a**). Generally, the output optical spectral bandwidth and average output power do not change much even changing multiplication conditions. As the SMSR changes for different multiplication condition, the amplitude modulation depth of the time-domain pulse train changes. Naturally, higher SMSR leads to lower amplitude modulation depth (e.g., 40 dB and 20 dB SMSRs lead to ~4% and ~40% amplitude modulation depth, respectively).

**Finding optimal injection conditions.** Several efforts are applied to effectively initiate injection locking and maximize the SMSR. First, a bandpass filter is employed to match the optical spectra between the injected seed signal and the mode-locked slave laser output (see Fig. 2). This is because only the mode-locking spectrum regime of slave laser can participate the injection locking. Second, the injection ratio is adjusted using variable optical attenuator. On one hand, lower injection ratio is desired to boost main modes and suppress side modes for higher SMSR. On the other hand, too low injection ratio can lead to the oscillation of the slave laser itself. Thus, we can maximally suppress the amplified side modes by properly setting injection power. Generally, one needs to use high input injection power to find a new injection locking condition. Once the injection locking is obtained, lowering injection power can lead to higher SMSR. Figure 9 shows the measured SMSRs for different injection input power and multiplication factors when a 250-MHz repetition-rate laser and a 73 – 78 MHz tunable repetition-rate laser are used as the master and slave oscillators, respectively. Note that the slave laser is pumped at 330 mW for all cases and the resulting output power was almost constant at ~1.1 mW with injection power at ~8 mW. As can be seen from Fig. 9, lowering input injection power (injection ratio) leads to higher SMSR: injecting less than 10 mW could lead to high SMSR performances.

## Acknowledgements

This research was supported by the National Research Foundation of Korea (Grant 2018R1A2B3001793)**.**

## Author contributions

C.-G.J., J.S. and J.K. designed the experiment. C.-G.J. and S.Z. performed the experiment and measured data. C.-G.J. and J.K. analysed the data and wrote the manuscript.

## Additional information

The authors declare no competing interests.

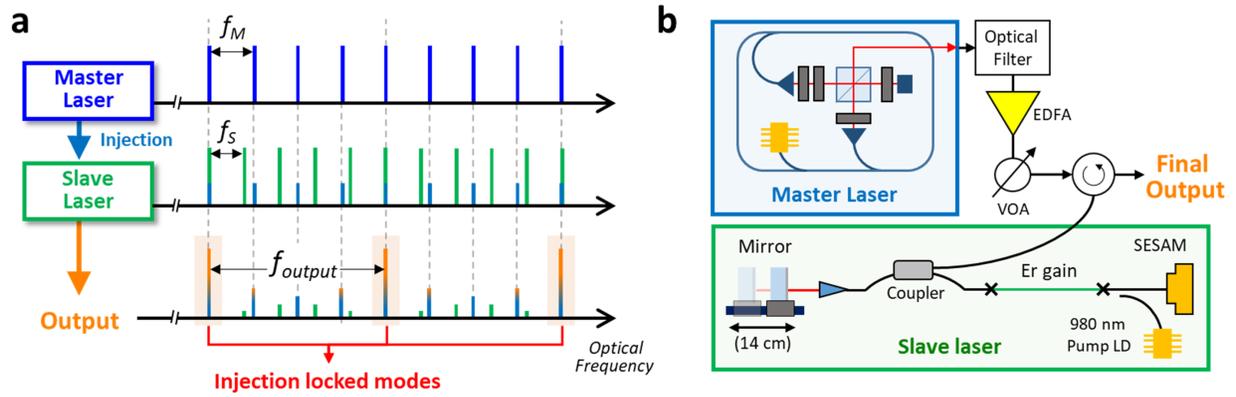

**Figure 1. Harmonic injection locking-based repetition-rate multiplication system. a** Conceptual principle of harmonic injection locking. $f_M$, repetition-rate of master laser; $f_S$, repetition-rate of slave laser; $f_{output}$, multiplied repetition-rate of the final output. **b** Overall schematic of the experimental setup. EDFA, Erbium-doped fibre amplifier; VOA, variable optical attenuator; SESAM, semiconductor saturable absorber mirror.

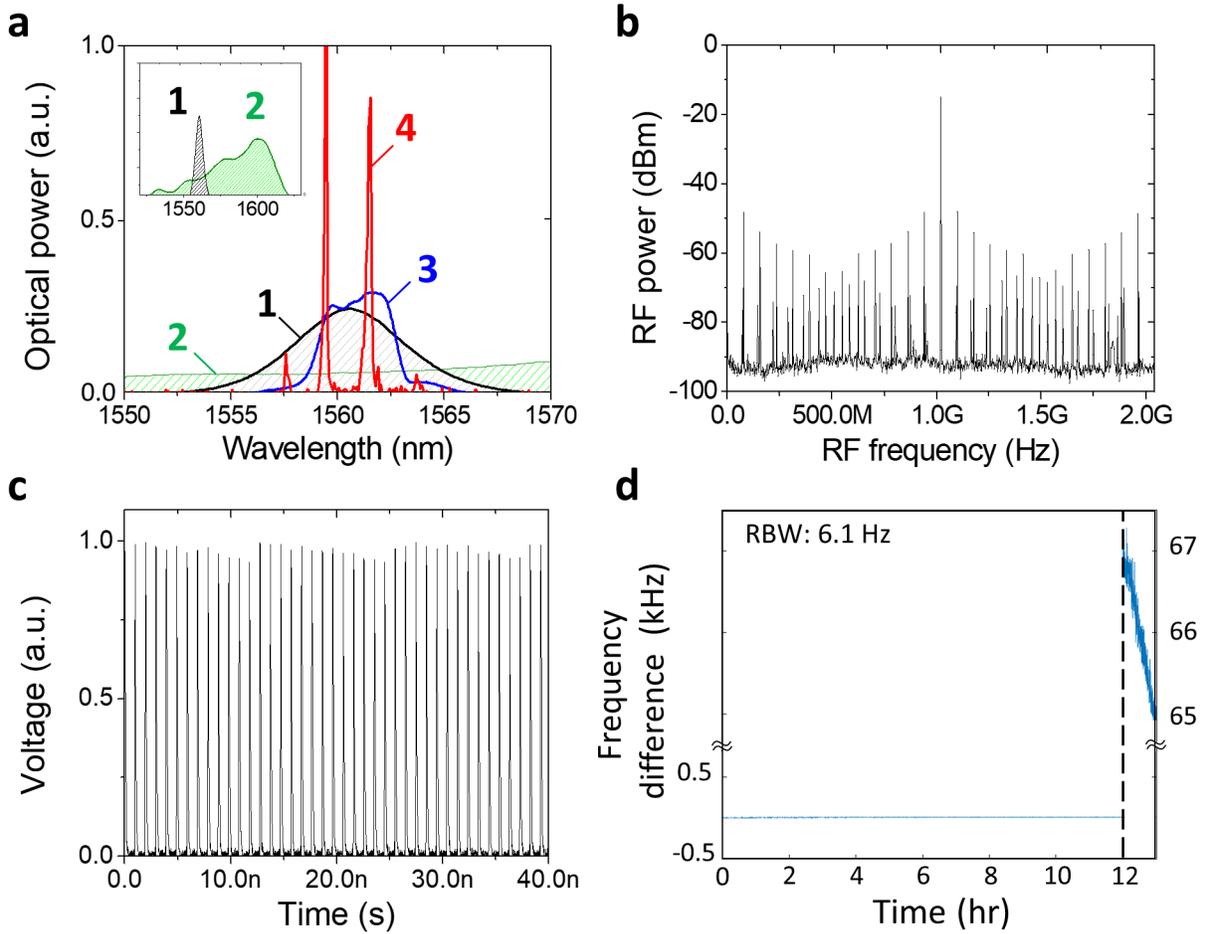

**Figure 2. Measurement results for RRM. (a)** Optical spectra of signals. The spectra 1 and 2 show the mode-locked spectrum of slave and master laser for each (broad range at inset), 3 is filtered master signal for injection and 4 is the final multiplied output of the RRM system. **(b)** RF spectrum of the repetition-rate multiplied output. **(c)** Time-domain measurement result of photodetected output signals. **(d)** Long-term measurement of relative frequency difference at the finial multiplied output frequency between master and slave lasers.

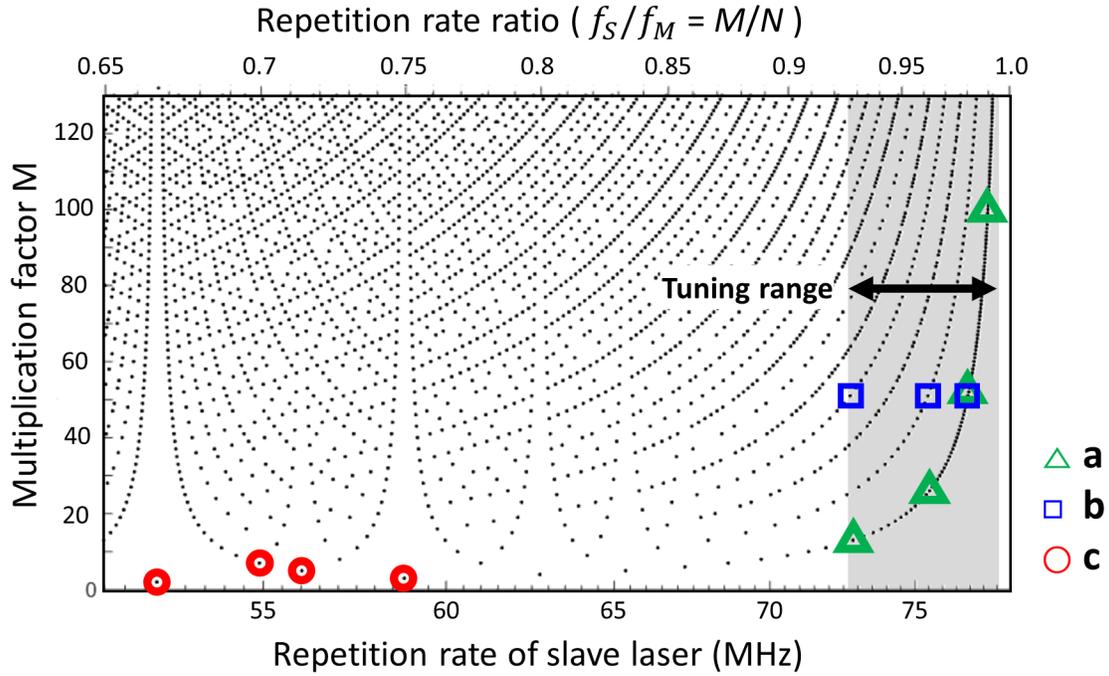

**Figure 3. Possible harmonic injection locking conditions for RRM.** (**a**) $N = (M+1)$ RRM conditions. $M = 13, 27, 51$ and $102$ (1 GHz, 2 GHz, 4 GHz, and 8 GHz, respectively) outputs are characterized for representative cases. (**b**) $M = 51$ RRM conditions with different slave oscillator conditions ($N = 52, 53$ and $55$). (**c**) $M = 2, 3, 5$ and $7$ RRM conditions. Note that the RF spectrum measurement results shown in Fig. 4**a**, 4**b** and 4**c** correspond to the regions (**a**), (**b**) and (**c**), respectively, in this figure.

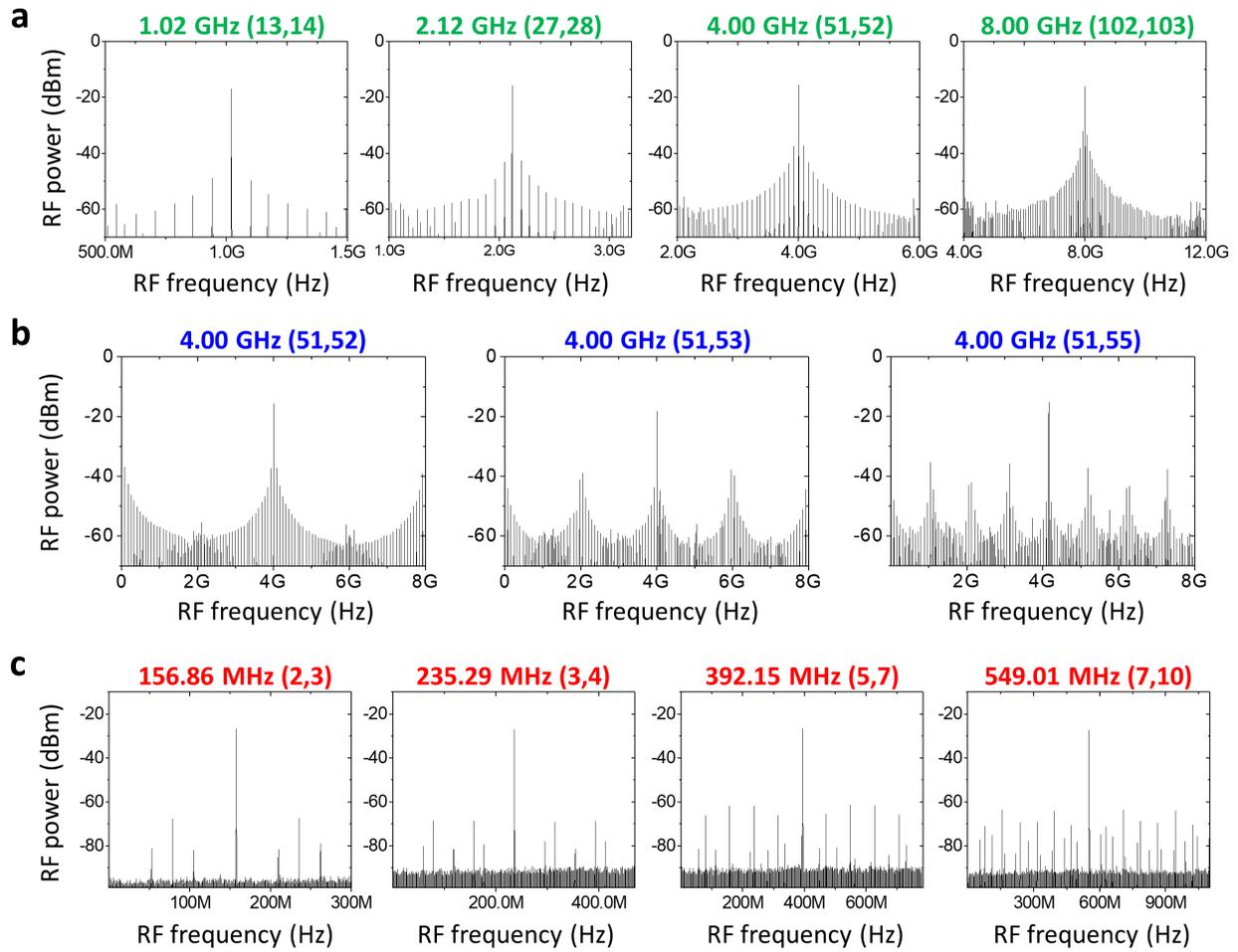

**Figure 4. RF spectra of the repetition-rate multiplied signals in various multiplication conditions.** (**a**) (*M,N*)=(13,14), (27,28), (51,52) and (102,103). (**b**) (*M,N*)=(51,52), (51,53), (51,55). (**c**) (*M,N*)=(2,3), (3,4), (5,7), (7,10).

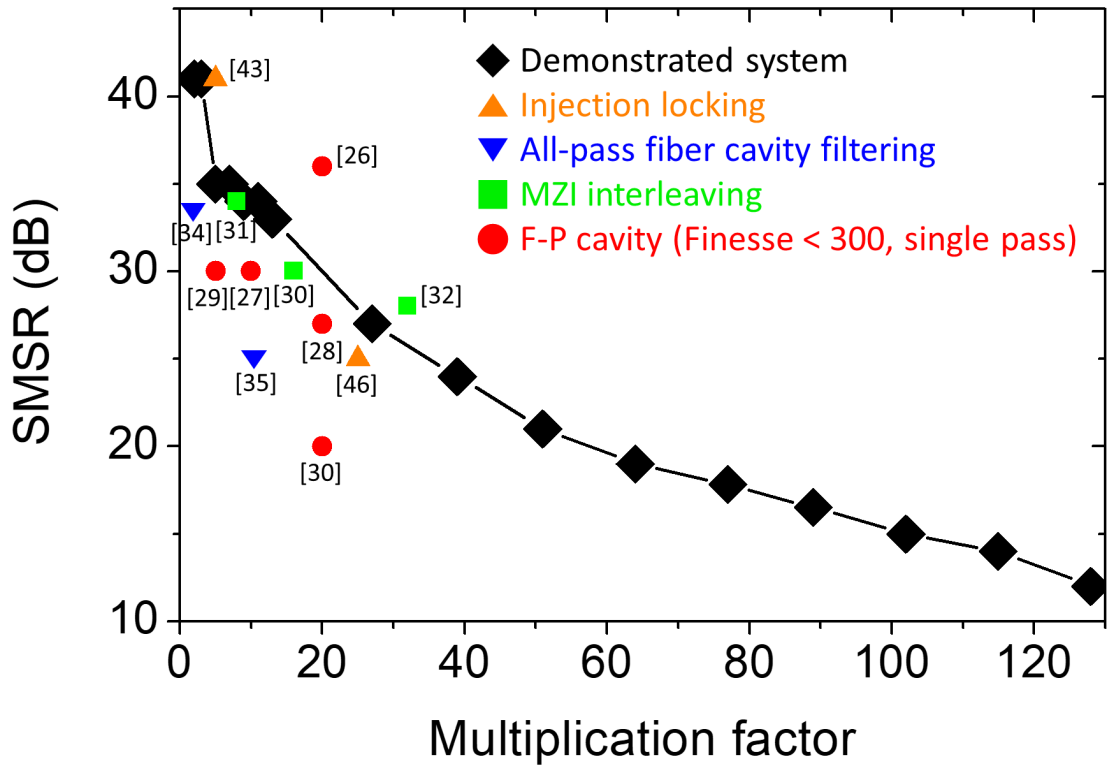

**Figure 5. SMSR variation versus multiplication factor and comparison with previous results.** Demonstrated system (black diamond) and related RRM methods using injection locking (orange triangle), all-pass cavity filtering (blue reversed triangle), MZI interleaving (green square), F-P cavity of finesse < 300 in single pass scheme (red circle).

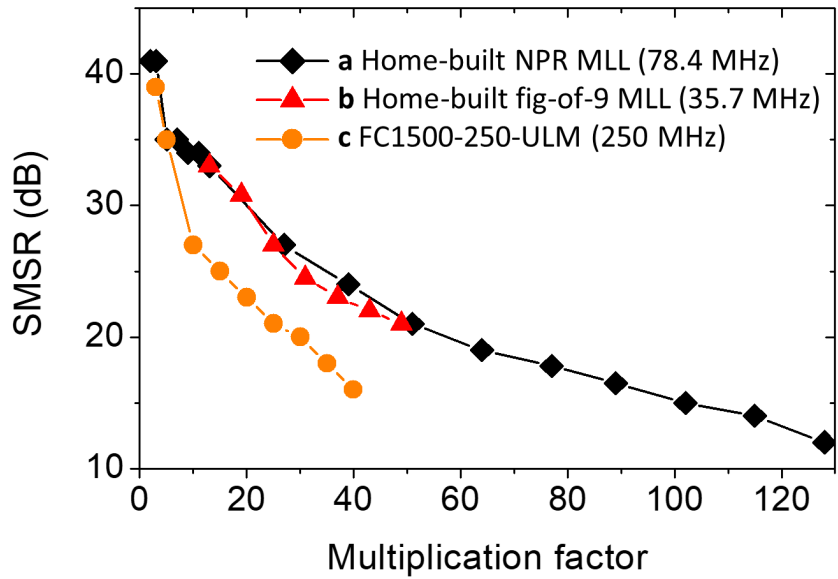

**Figure 6. SMSR variation versus multiplication factor for different master lasers.** (**a**) 78.43-MHz home-built nonlinear polarization rotation-based mode-locked fibre laser. (**b**) 35.7-MHz home-built nonlinear amplifying loop mirror-based mode-locked fibre laser. (**c**) 250-MHz nonlinear amplifying loop mirror-based mode-locked fibre laser (MenloSystem GmbH, FC1500-250-ULN)

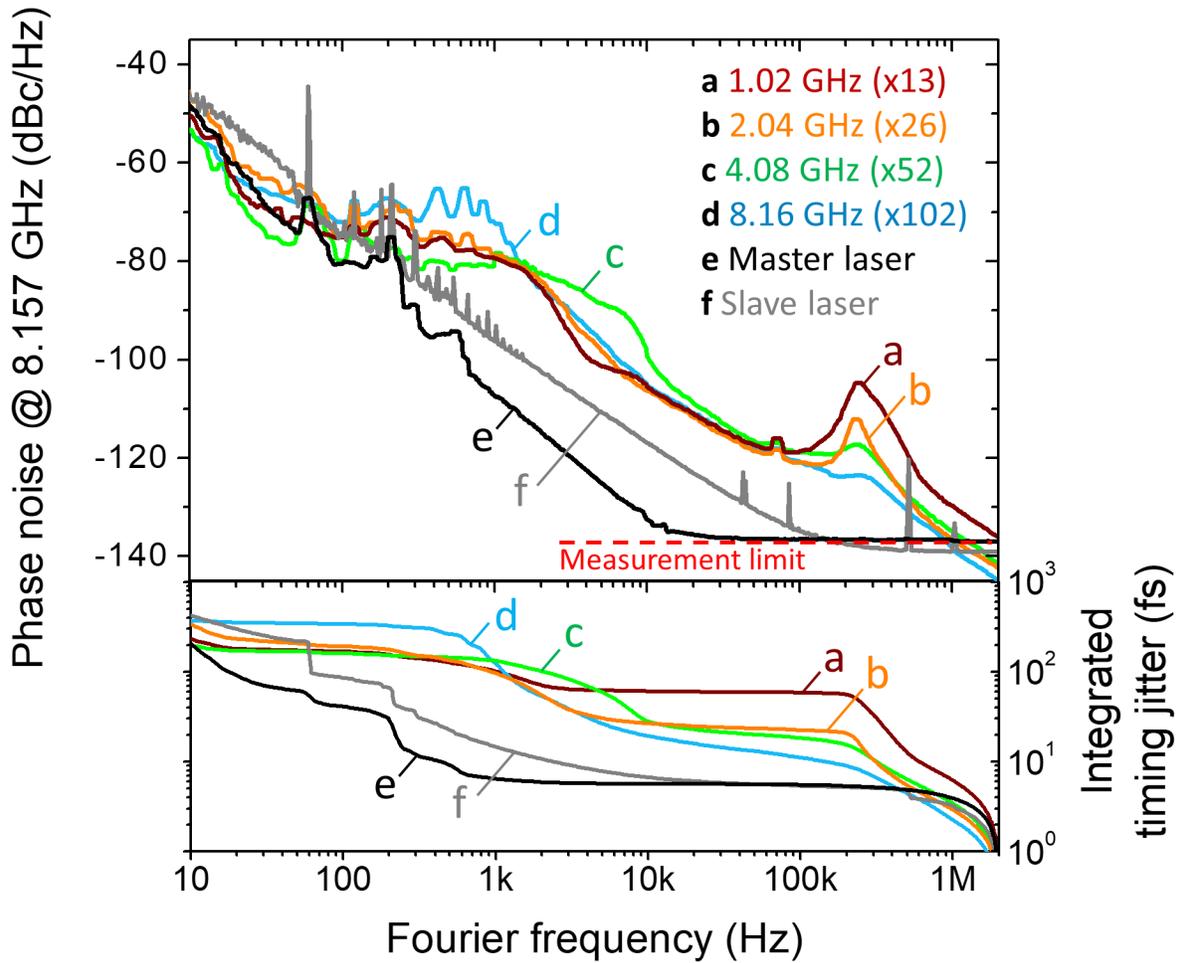

**Figure 7. Repetition-rate phase noise and integrated timing jitter of RMM output signal.** (**a**) *M*=13 (1.02 GHz). (**b**) *M*=26 (2.04 GHz). (**c**) *M*=52 (4.08 GHz). (**d**) *M*=102 (8.16 GHz). (**e**) 78.34-MHz master laser. (**f**) 72.84-MHz slave laser.

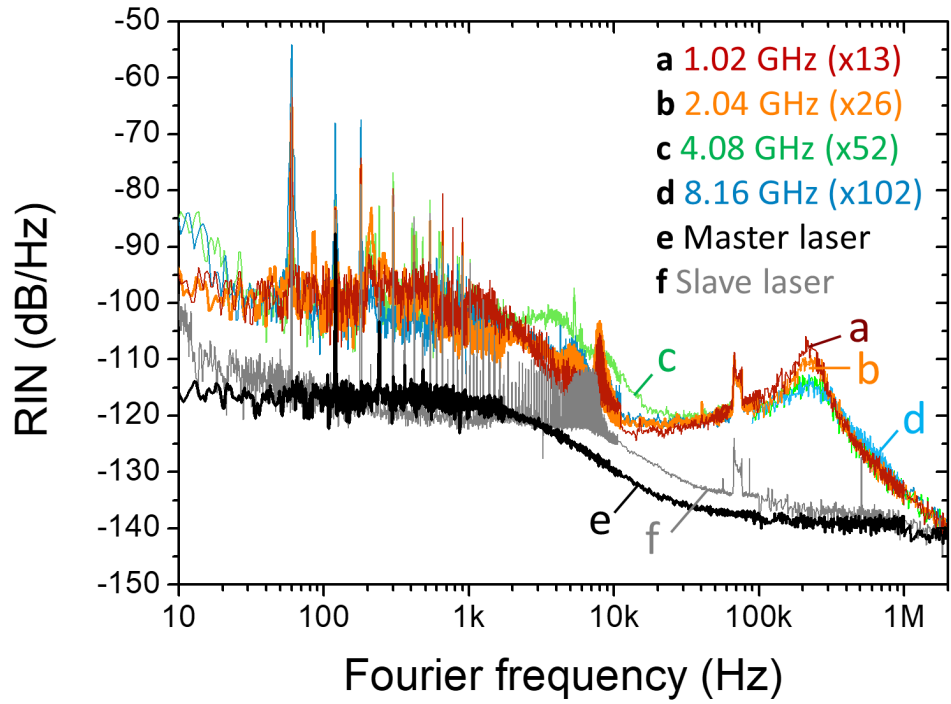

**Figure 8. Relative intensity noise (RIN) of RMM output signal.** (**a**) *M*=13 (1.02 GHz). (**b**) *M*=26 (2.04 GHz). (**c**) *M*=52 (4.08 GHz). (**d**) *M*=102 (8.16 GHz). (**e**) 78.34-MHz master laser. (**f**) 72.84-MHz slave laser.

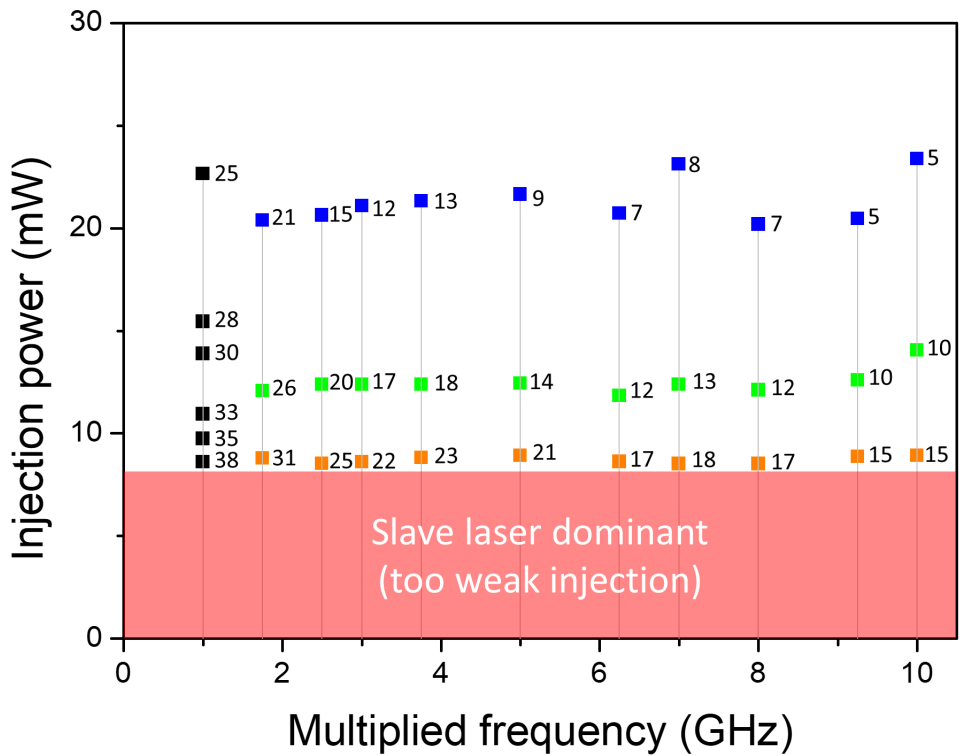

**Figure 9. Injected power versus SMSR for different multiplication factors.** Each number indicates the measured SMSR in dB unit. A 250-MHz master laser and a 73-78 MHz tunable slave laser are used for the experiment.